\documentclass[epj]{svjour}

\usepackage{graphicx}
\usepackage{fancyhdr}
\usepackage{amssymb}

\def\mytitle{Stability and Leptogenesis in Left-Right Symmetric
Seesaw Models} \def\myauthors{Tomas H\"allgren}
\def\mytype{Contributed Talk} \def\mysession{Cosmology}

\begin{document}
\title{Stability and Leptogenesis in Left-Right Symmetric Seesaw Models}
\subtitle{}
\author{Tomas H\"allgren\inst{}\thanks{\emph{Email:} tomashal@kth.se}%
\,\thanks{This talk is based on Ref.~\cite{Hallgren1}.}%
}                     
%
%
\institute{Department of Theoretical Physics, School of Engineering Sciences\\
Royal Institute of Technology (KTH)\\
AlbaNova University Center\\
Roslagstullsbacken 21, 106 91 Stockholm, Sweden
}
%
\date{}
\abstract{In left-right symmetric seesaw models an eight-fold
degeneracy among the right-handed neutrino mass matrices is known to
exist. We use stability and viability of leptogenesis as criteria in
order to discriminate among the degenerate solutions and to partially
lift the eight-fold degeneracy.
\PACS{{98.80.Cq}{Particle-theory and field-theory models of the early Universe}\and
      {14.60.Pq}{Neutrino mass and mixing}\and{14.60.St}{Non-standard-model neutrinos, right-handed neutrinos}
     } 
} 
\maketitle
\section{Introduction}
It has become an established fact that neutrinos are massive, although
very light. The Standard Model (SM) of particle physics is
insufficient in explaining in a natural way why the neutrinos are so
much lighter than the charged leptons. One attractive and elegant
solution to this problem is provided by the seesaw mechanism
\cite{seesaw}, where the small masses of the neutrinos are induced by
the presence of heavy particles, such as right-handed Majorana
neutrinos or Higgs triplets. The seesaw mechanism has in addition the
attractive feature of a built-in mechanism for generating the baryon
asymmetry of the Universe (BAU) through the baryogenesis via
leptogenesis mechanism \cite{leptogenesis}, whereby the decays of the
right-handed neutrinos produce a lepton asymmetry, which is then
partly converted into a baryon asymmetry by sphaleron processes.

Recently, it was demonstrated \cite{Frigerio,Frigerio2} that in a
certain class of left-right symmetric seesaw models, the seesaw mass
formula can be inverted to yield eight (in the three-flavor case)
possible mass matrices for the right-handed Majorana neutrinos, for a
given light neutrino mass matrix and Dirac Yukawa couplings as
input. Since the right-handed neutrino sector is not directly
accessible by laboratory experiments, it is important to find ways of
discriminating among the eight seesaw solutions. This was the
objective of Refs.\cite{Hallgren1,Lavignac}, where leptogenesis and
stability of the solutions were used as selection criteria and it was
shown that it is possible to partially lift the degeneracy in the
right-handed neutrino sector.
\begin{figure*}
\includegraphics[clip,width=0.9\textwidth,height=0.35\textwidth,angle=0]{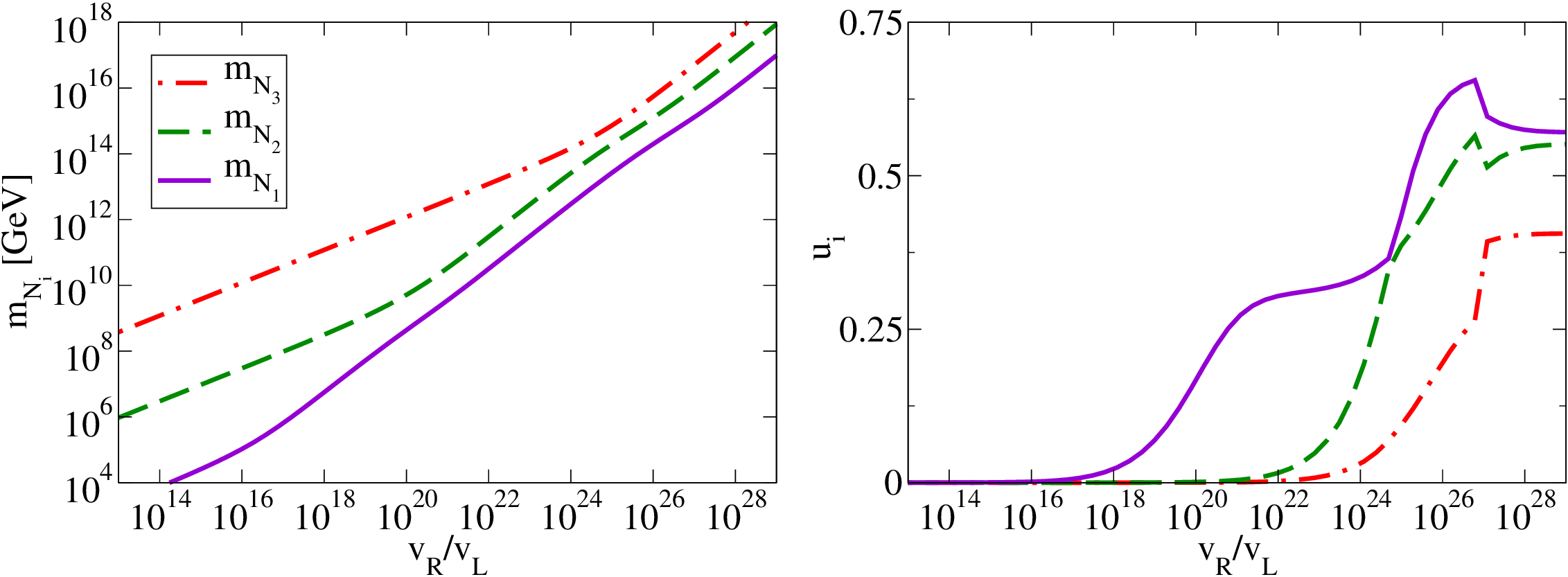}
\caption{The right-handed neutrino masses $m_{N_i}$ and mixing
parameters $u_i$ as functions of $v_R/v_L$ for the solution
'$+++$'. Inverted mass hierarchy, $m_0=0.1$~eV. Figure from
Ref.~\cite{Hallgren1}.}
\label{fig:fig1}       
\end{figure*}
\section{The left-right symmetric seesaw model}
We will consider the left-right symmetric seesaw model with gauge
group $G = SU(2)_{L}\times SU(2)_{R}\times U(1)_{B-L}$
\cite{Pati,Mohapatra}. The SM field content is extended to include three heavy
right-handed Majorana neutrinos $N_{i}$ and the Higgs sector contains
an $SU(2)_{L}\times SU(2)_{R}$ bidoublet $\phi$ as well as
$SU(2)_{L,R}$ triplets $\Delta_{L,R}$. When the neutral components of
the Higgs fields acquire vacuum expectation values (VEVs) we obtain
the type I+II seesaw mass formula
\begin{equation}\label{eq:seesaw}
m_\nu = f \, v_L - \frac{v^2}{v_R} \, y \, f^{-1} y^T\,, 
\label{eq:ss1}
\end{equation}
where $v, v_L,$ and $v_R$ denote the VEVs of the neutral components of
 $\phi$ and $\Delta_{L,R}$, $y$ denotes the Dirac Yukawa coupling
 matrix and $f$ the triplet Yukawa coupling matrix. In general, the
 triplet Yukawa couplings for $\Delta_{L,R}$ could be different, but
 we will restrict to the case where there is a discrete left-right
 symmetry such that $f_{L}=f_{R}\equiv f$ and $y$ is symmetric. This
 case has much fewer parameters and is more predictive. We will assume
 that the Dirac Yukawa coupling matrix $y$ equals the up-type quark
 Yukawa coupling matrix $y_{u}$, which is motivated by grand unified
 theories (GUTs). Many of the results depend mostly on the fact that
 $y$ has hierarchical eigenvalues. The free parameters will then be
 the absolute neutrino mass scale $m_{0}$, the ratio of the triplet
 VEVs, and the light neutrino mass hierarchy. The seesaw formula
 Eq.~(\ref{eq:seesaw}) can now be inverted to yield $2^{N}$ solutions
 (for $N$ generations) for $f$, and hence $2^{N}$ possible
 right-handed Majorana mass matrices, which are given by
 $M_{R}=fv_{R}$.  In the one-generation case, the two solutions admit
 a simple analytical form
\begin{equation}
f_{\pm} = \frac{m_\nu}{2v_L} \pm \frac{1}{v_L}\sqrt{\frac{m^2_\nu}{4} + \frac{y^2 v_L v^2}{v_R}}.
\end{equation}
Analytical solutions exist also in the three-flavor case and we label
the eight solutions as '$\pm\pm\pm$' where '$-$'('$+$') denotes that
the corresponding mass eigenvalue is type I (II) dominated in the
large $v_{R}/v_{L}$ regime. The labeling starts with the largest
eigenvalue in the small $v_{R}/v_{L}$ regime. Thus, there will be two
solutions which are either pure type I or type II dominated for large
$v_{R}/v_{L}$ and six mixed solutions.

In Fig.~\ref{fig:fig1} we show the eigenvalues and mixing for the
solution '$+++$' for $m_{0}=0.1~$eV and inverted light neutrino mass
hierarchy. As a measure of mixing, we have introduced the parameters
$u_i$ which are related to the off-diagonal elements of the unitary
matrix $U$ which diagonalize $f$ as:
\begin{eqnarray}
u_1^2 & = &  \frac12 (|U_{12}|^2 + |U_{21}|^2)\,, \quad 
u_2^2 = \frac12 (|U_{13}|^2 + |U_{31}|^2),\nonumber\\&&u_3^2 = \frac12 (|U_{23}|^2 + |U_{32}|^2).
\end{eqnarray}
In the next section, we will consider ways of selecting among the
eight solutions.
%
%
%



\section{Discriminating among the degenerate solutions}
\subsection{Stability}
\begin{figure}
\includegraphics[clip,width=0.45\textwidth,height=0.32\textwidth,angle=0]{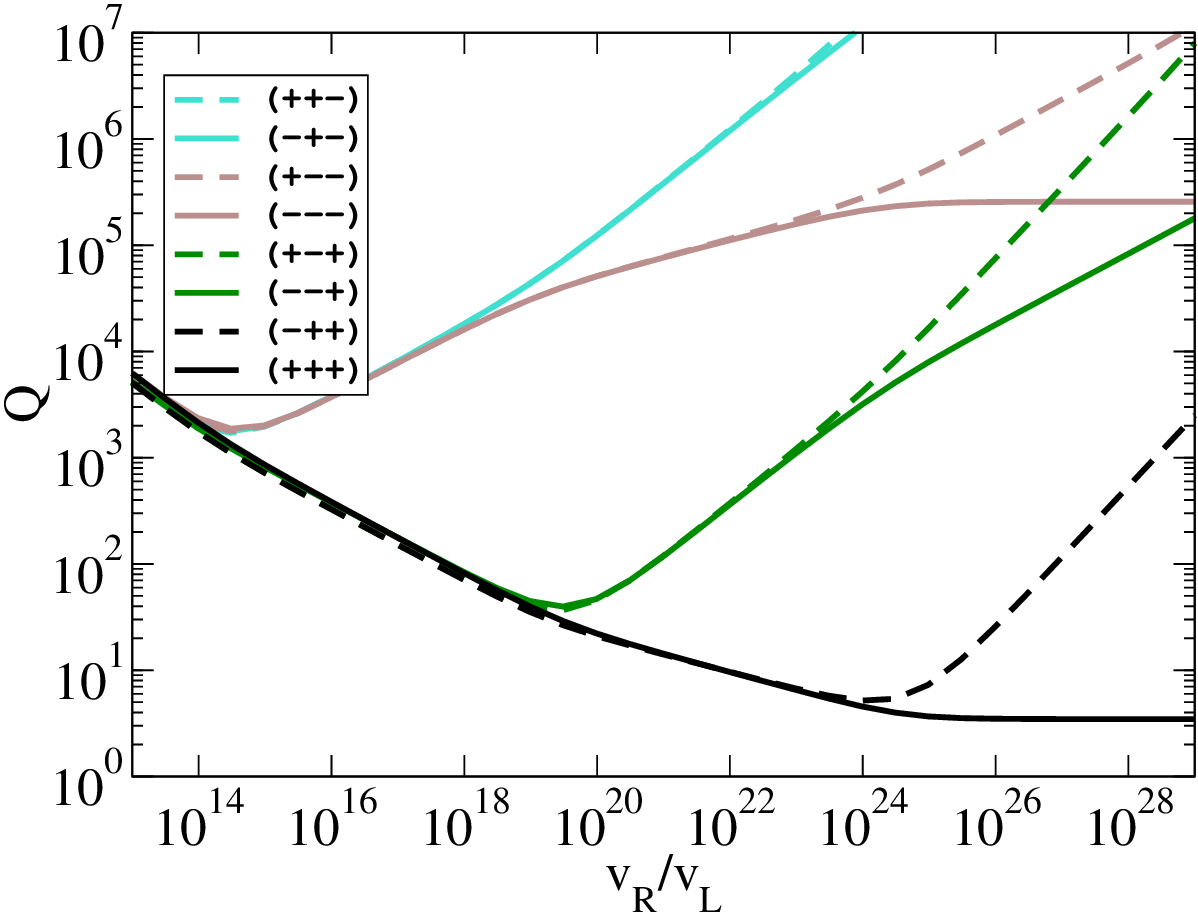}
\caption{The stability measure $Q$ as a function of $v_R/v_L$ for
$m_0=0.1$ eV and inverted hierarchy. Figure from
Ref.~\cite{Hallgren1}.}
\label{fig:fig2}       
\end{figure}

First we will introduce a notion of naturalness, or stability, in
order to select among the degenerate solutions. We pose the question
if, for a given light neutrino mass matrix, the triplet Yukawa
coupling matrix $f$ has to be very special or fine-tuned in the sense
that a marginally different $f$ would give very different low-energy
neutrino phenomenology. We consider such a situation unnatural. In
order to quantify the stability of the solutions we introduce the
following stability measure
\begin{equation}
\label{eq:Qmeasure}
Q = \left| \frac{\det{f}}{\det{m_\nu}} \right|^{1/3} \sqrt{\sum_{k,l=1}^{2N} 
\left( \frac{\partial {m_l}}{\partial f_k} \right)^2}\,.
\end{equation}
The matrices $f$ and $m_\nu$ are determined by the real coefficients
$f_k$ and $m_l$ according to
\begin{equation}
f = \sum_k ( f_k + i f_{k+N}) T_k,
m_\nu = \sum_k ( m_k + i m_{k+N}) T_k, 
\end{equation}
where $T_k$, $k\in[1,6]$, form a normalized basis of complex symmetric
$3\times 3$ matrices. This stability measure can be shown to be basis
independent \cite{Hallgren1}. Fig.~\ref{fig:fig2} shows the stability
measure $Q$ for the eight solutions, for $m_0 = 0.1$ eV and an
inverted light neutrino mass hierarchy. First we note that all
solutions are unstable for small $v_{R}/v_{L}$, which is due to the
fact that in this regime there must be a very precise cancellation
between the type I and type II contributions to the light neutrino
mass matrix in Eq.~(\ref{eq:seesaw}). In addition, most of the solutions
also become unstable for large $v_{R}/v_{L}$. This happens when there
is a large spread in the eigenvalues, which imply suppressed mixing
and fine-tuning between $f$ and $y$ (for a detailed discussion see
Ref.~\cite{Hallgren1}). The most stable solution is the purely type II
dominated '$+++$' solution. If one allows for a fine-tuning at the
percent level, $Q\lesssim 10^{3}$, then the two solution '$\pm++$',
for $v_{R}/v_{L}\gtrsim 10^{18}$ and the solutions '$\pm-+$' with
$v_{R}/v_{L}\simeq 10^{20}$ are favored. The stability results do not
change qualitatively, when changing $m_{0}$ or the light neutrino mass
hierarchy. Additional CP violating phases only change the results
marginally.
%
\begin{figure}
\includegraphics[clip,width=0.45\textwidth,height=0.3\textwidth,angle=0]{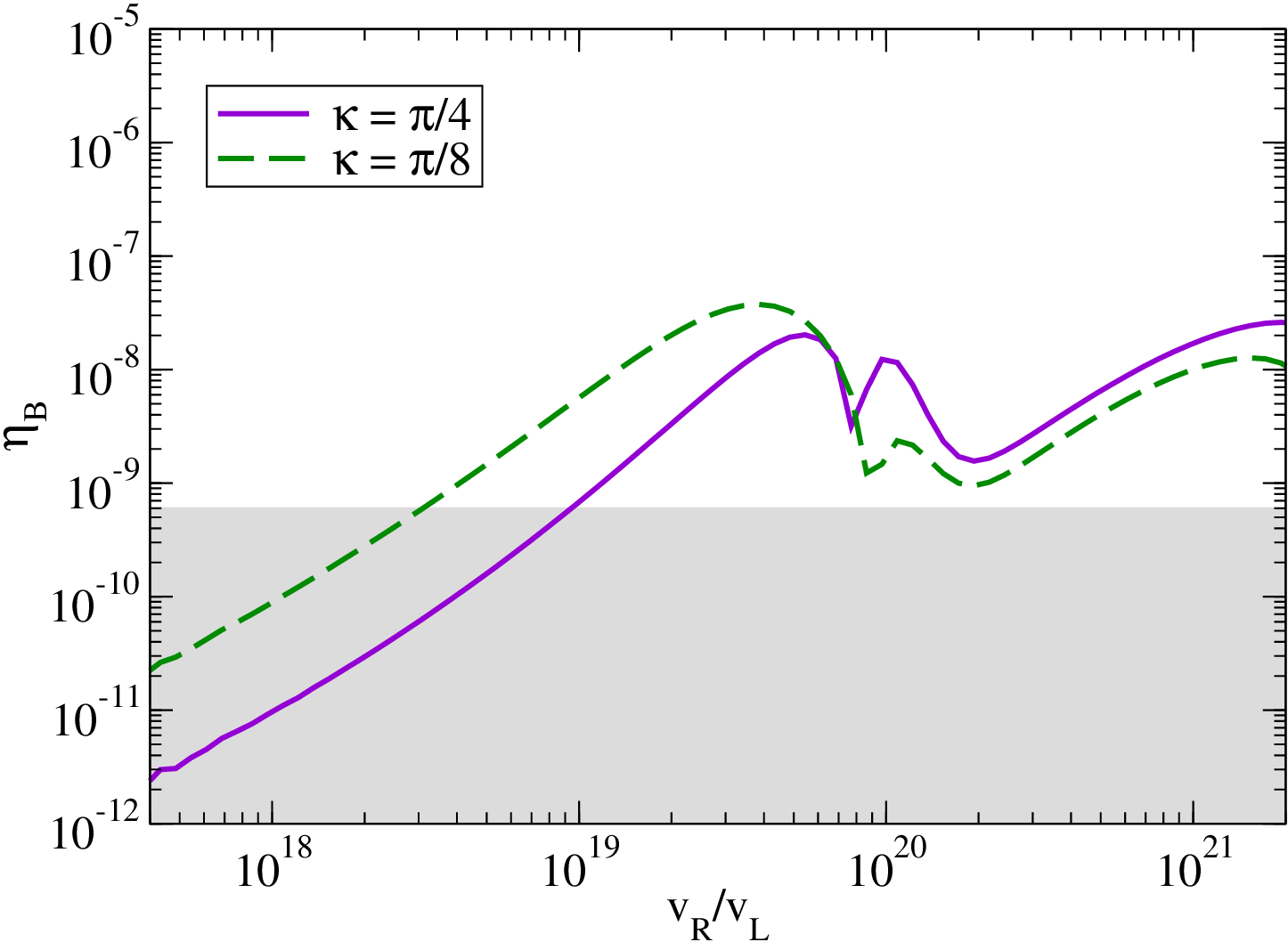}
\caption{The baryon-to-photon ratio $\eta_B$ with an additional complex phase
$\pi/8$ or $\pi/4$ attributed to the electron neutrino for the
solution '$+++$'. The shaded area corresponds to values of $\eta_B$
below the observed value. Inverted mass hierarchy, $m_0 = 0.1$~eV. Figure from Ref.~\cite{Hallgren1}.}
\label{fig:fig3}   
\end{figure}
\subsection{Leptogenesis}
The second selection criteria we use is viability of leptogenesis for
a given solution. Thus, we ask whether a solution can reproduce the
observed BAU, $\eta_B = (6.1 \pm 0.2) \times 10^{-10}$
\cite{Spergel:2003cb}. The baryon asymmetry can be parametrized as
\begin{equation}\label{eq:etaB}
\eta_B \equiv \frac{n_B}{n_\gamma} = \eta \, \epsilon_{N_1},
\label{eta_fac}
\end{equation}
where $\eta$ is the so-called efficiency factor that takes into
account washout effects, the initial density of the right-handed
neutrinos and the deviation from equilibrium in their decays and
$\epsilon_{N_1}$ denotes the CP-asymmetry parameter, which measures
the CP-asymmetry in the decay of the lightest right-handed
neutrino. In Eq.~(\ref{eq:etaB}) we have assumed that the lightest
right-handed neutrino is much lighter than the other heavy neutrinos,
such that it gives the dominating contribution to the lepton
asymmetry. In Ref.~\cite{Hallgren2} we study leptogenesis in the same
framework, but where instead the decay of the triplet $\Delta_L$ is
the source of the lepton asymmetry. The CP-asymmetry parameter is
defined as
\begin{equation}
\epsilon_{N_i} = \frac{\Gamma(N_i \to l\, H) - \Gamma(N_i \to \bar l\, H^*)}
	{\Gamma(N_i \to l\, H) + \Gamma(N_i \to \bar l\, H^*)}\,.
\end{equation}
From a numerical solution of the Boltzmann equations, it is possible
to derive the following expression for the efficiency parameter
\begin{equation}
\eta = 1.45 \times 10^{-2} \frac{10^{-3} \, {\rm eV}}{\tilde m_1}, 
\end{equation}
assuming a mass hierarchy in the right-handed neutrino sector. Here we
have also introduced the effective neutrino mass
\begin{equation}
\tilde m_i = \frac{v^2 \, (\hat y^\dagger \hat y)_{ii}}{2m_{N_i}}\,.
\end{equation}
The notation with a hat indicates that all matrices are evaluated in
the basis where $f$ is real and diagonal. The CP-asymmetry parameter
receives contributions of type I (from the right-handed neutrinos) and
type II (from the triplets). In the limit where the lightest
right-handed neutrino is considerably lighter than the heavier
neutrinos as well as the triplets, the CP-asymmetry can be written as
\cite{Antusch}
\begin{equation}
\epsilon_{N_1} = \epsilon^I_{N_1} + \epsilon^{II}_{N_1} \to \frac{3}{16 
\pi} \frac{\hat f_{11}v_R}{v^2} \frac{{\rm Im}[(\hat y^\dagger \, \hat m_\nu \hat y^*)_{11}]}
{(\hat y^\dagger \hat y)_{11}}\,.
\label{lep_approx}
\end{equation}
Our numerical results indicate that the upper bound on the CP
asymmetry found in Ref.~\cite{Antusch} can be saturated. An important
feature of the left-right symmetric model is the presence of
additional physical Majorana phases, which can increase the CP
asymmetry and improve the prospects for leptogenesis. As demonstrated
in Ref.~\cite{Hallgren1}, this allows for leptogenesis even with only
one right-handed neutrino. In the case of three generations there are
more sources of CP violation, although mixing can increase the
effective neutrino mass, thereby increasing the
washout. Fig.~\ref{fig:fig3} shows that baryon-to-photon ratio
$\eta_{B}$ for the solution '$+++$', when a Majorana phase $\kappa$
is attributed to the electron neutrino. We have chosen $\kappa
=\pi/4\,(\pi/8)$. The mass of the lightest right-handed neutrino
required to reproduce the observed BAU is $m_{N_{1}}\gtrsim 1.4\times
10^{9}~$GeV\\$(m_{N_1}\gtrsim 2.5\times 10^8~$GeV). We find that
leptogenesis is possible for four out of the eight solutions. For the
other solutions, leptogenesis is not possible, since for these cases
the mass of the lightest right-handed neutrino never exceeds
$10^{6}~$GeV. It should be noted that the results depend on the
assumption that $y=y_{u}$. For a different choice of Yukawa couplings
it is possible to relax the mass bounds.  However, the choice
$y=y_{u}$, as discussed before, is motivated by GUTs. Our results
complements the results of the leptogenesis analysis in
Ref.~\cite{Lavignac}, since the washout is less severe for our
specific choice of parameters.
\begin{table*}
\begin{center}
\begin{tabular}{llll}
\hline\noalign{\smallskip}
 & $\pm++$ & $\pm-+$ & $\pm\pm-$ \\
\noalign{\smallskip}\hline\noalign{\smallskip}
Stability & $v_R/v_L > 10^{18}$ & $v_R/v_L \simeq 10^{20}$ &  disfavored \\
Leptogenesis & $v_R/v_L > 10^{18}$ & $v_R/v_L > 10^{18}$ &
excluded\\
Gravitinos &  $v_R/v_L < 10^{21}$ & unconstrained
& unconstrained\\
\noalign{\smallskip}\hline
\end{tabular}
\end{center}
\caption{The allowed regions of the parameter $v_R/v_L$ for the eight
different types of solutions. Table from Ref.~\cite{Hallgren1}.\label{finaltab}}
\end{table*}
\section{Summary}
We have studied the left-right symmetric type I+II seesaw mechanism
with a hierarchical Dirac Yukawa coupling matrix motivated by GUTs.
It has been shown previously that it is possible to invert the seesaw
mass formula to obtain eight possible triplet Yukawa coupling matrices
for a light neutrino mass matrix and Dirac Yukawa coupling matrix as
input. Our goal was to discriminate among these degenerate solutions
using stability properties and viability of leptogenesis as
criteria. As a measure of stability we have introduced the parameter
$Q$ (see Eq.~(\ref{eq:Qmeasure})) which quantifies the amount of
fine-tuning.

Our results are summarized in Tab.~\ref{finaltab}. The stability
criterion disfavors the four solutions of the type '$\pm\pm-$' and
favors the solutions '$\pm-+$' provided that $v_R/v_L \simeq
10^{20}$. The remaining two solutions '$\pm++$' are stable, for
$v_R/v_L \gtrsim 10^{18}$. In addition, we find that successful
leptogenesis is possible for the four solution of the type '$\pm\pm+$'
if $v_R/v_L \gtrsim 10^{18}$. This requires the existence of
additional Majorana-type phases, not present in the pure type I seesaw
framework. Additional constraints would also be present in a
supersymmetrized version, coming from the potentially dangerous
overproduction of gravitinos in the early universe. For our specific
choice of Yukawa couplings, $y=y_{u}$, only the solutions of the type
'$\pm++$' are affected by this constraint. This leads to the
requirement $v_R/v_L \lesssim 10^{21}$.\\ Thus, we have shown, that it
is possible, within the chosen framework, to partially lift the
eight-fold degeneracy among the right-handed neutrino mass matrices in
the left-right symmetric seesaw model.

\end{document}